\documentclass[%
 reprint,
 amsmath,amssymb,
 aps,
]{revtex4-2}

\usepackage{graphicx}
\usepackage{dcolumn}
\usepackage{bm}

\begin{document}

\preprint{arxiv}

\title{Velocity distributions of particles sputtered from supported 2D-MoS$_2$ \\ during highly charged ion irradiation}

\author{Lucia Skopinski,\textit{$^{a}$} Silvan Kretschmer,\textit{$^{b}$} Philipp Ernst,\textit{$^{a}$} Matthias Herder,\textit{$^{a}$} Lukas Madauß,\textit{$^{a}$}  Lars Breuer,\textit{$^{a}$} Arkady V. Krasheninnikov$^{\ast}$ \textit{$^{b,c}$}, and Marika Schleberger$^{\ast} $\textit{$^{a}$}}
\affiliation{$^a$Fakultät für Physik and CENIDE, Universität Duisburg-Essen, 47057 Duisburg, Germany}
\affiliation{$^b$Institute of Ion Beam Physics and Materials Research, Helmholtz-Zentrum Dresden-Rossendorf, 01328 Dresden, Germany}
\affiliation{$^{c}$Department of Applied Physics, Aalto University, 00076 Aalto, Finland}

\date{\today}
\date{\today}

\begin{abstract}
The interaction of highly charged ions (HCI) with solids leads to particle sputtering, which can be used for defect-mediated engineering of the properties of the material. Ions can store energy in the form of kinetic and potential energy (sum of the ionization energies) and transfer it to the solid upon impact. The interaction and sputtering mechanisms depend significantly on the projectile energies. However, the relevance of various interaction mechanisms is unknown. Here we show that for slow HCI (5 keV) the interaction mechanisms leading to particle emission by electronic excitation and transferred kinetic energy are independent from each other, which is consistent with our atomistic simulations. We have irradiated substrate supported (Au, SiO$_2$) monolayers of MoS$_{2}$ with highly charged xenon ions (charge state: 17+ -- 40+), extracted the emitted neutral, post-ionized Mo particles into a time-of-flight mass spectrometer and determined their velocity distributions. We find two main contributions, one at high velocities and a second one at lower velocities, and assign them to kinetic and potential effects respectively. Our data suggests that the dominant mechanism for potential sputtering is related to electron-phonon coupling, while non-thermal processes play no significant role. We anticipate that our work will be a starting point for further experiments and simulations to determine whether the different processes resulting from E$_{pot}$ and E$_{kin}$ can be separated or whether synergistic effects play a role. 
\end{abstract}

\maketitle

\section{Introduction}
Two-dimensional (2D) materials are expected to be the key elements of many novel devices due to their unique mechanical, electronic, and chemical properties. After the first preparation of graphene in 2004,\cite{Novoselov.2004} the number of realized 2D structures has increased to several hundred so far.\cite{Mounet.2018} Apart from improving the synthesis,\cite{Lee.2012, Lin.2016} a main focus in research lays on the processing and modification of the materials towards the realization of devices. One way to influence their properties is by stacking several monolayers and therefore creating van der Waals heterostructures with tuneable or synergistic properties.\cite{Geim.2013, Novoselov.2016} In addition, surface structuring methods, like beams of various energetic particles have been suggested and successfully applied for defect engineering.  \cite{Schleberger.2018, Madau.2018, Wang.2018}
While ion beams are an established method to structure thin films or 2D materials, with the latter several problems occur. First, the description by classical models is not straightforward. Sputtering and implantation, e.g., are different due to the absence of a 3D crystal lattice. Further, atoms sputtered from the substrate can cause severe damage to the 2D monolayer on top. \cite{Kretschmer.2018} This problem can be circumvented, e.g., by using slow highly charged ions (HCI) with low kinetic energy (in the keV range) and high potential energy (created by the sum of the ionization energies).\cite{Aumayr.2011} These projectiles deposit their energy into the electronic system of the target material near its surface. It has been shown that HCIs in particular can be applied as an efficient tool for pore creation and material modification. For example, pore creation in freestanding MoS$_2$ allows for adjustment of the pore radius by tuning the charge state of the primary HCI \cite{Kozubek.2019} and pore creation in insulating fluorinated graphene works equally well.\cite{Creutzburg.2021} Highly conductive, pristine graphene, on the other hand, has been proven to be very resistant as its high electron mobility prevents pore formation due to electronic excitation \cite{Gruber.2016} and may even work as a protective cover.\cite{Schwestka.2020} 

These examples show that the energy dissipation within the target material depends strongly on its electronic and thermal properties. However, the exact mechanisms that lead to surface structures are still a subject of discussion. Especially for semiconducting and insulating targets with lower electron mobility the high local perturbation in the electronic system caused by neutralization and de-excitation of the impinging HCI is not well understood. In fact, the proposed models range from Coulomb explosion over thermal spike and defect mediated desorption to non-thermal melting. Although these models differ greatly in terms of their spatial and temporal dynamics, the final state is often similar (some amount of target material is missing), making post-mortem analysis inadequate to distinguish between models. This poses a key problem, in particular if a specific modification of a 2D material, whether free-standing, substrate-supported, or in a heterostructure, is to be achieved. 

To shed light on the ion-solid interaction mechanisms, we therefore take another approach. By means of mass spectrometry we study particle emission from a supported 2D material {\it during} HCI irradiation. By varying both the charge state and kinetic energy of the ion as well as substrate, we can systematically exploit the parameter space to identify and evaluate the individual contributions. In addition, we study the velocity distribution of particles emitted by HCI irradiation and compare molecular dynamics (MD) simulations that have been performed to further our understanding of the experimental findings. Our work thus helps to establish guide-lines for defect engineering of 2D materials by kinetic and electronic excitation. 



\section{Materials and Methods}
\noindent
\textbf{Sample Preparation}\\
Monolayer MoS$_2$ flakes were grown directly on a 300 nm SiO$_2$/Si substrate by CVD. To this end, a 1 \% sodium
cholate solution as growth promoter is spin-coated at 4000 rpm for 60 s onto the cleaned (acetone, sonication) SiO$_2$ substrate. A droplet of a saturated solution of ammonium heptamolybdate in deionized (DI) water is deposited onto the substrate and then heated for 24 min at 300 $^o$C to form MoO$_3$, providing the molybdenum feedstock. The substrate is placed in the center of a 1 inch CVD tube furnace and 40 mg of solid sulphur (Sigma Aldrich; 99,89 \%) is placed 15 cm upstream from the substrate in a different heating zone. Growth occurs at atmospheric pressure in a flow of 500 sccm of Ar gas (99.999 \% purity). The furnace temperature is ramped to 750 $^o$C at a rate of 75 $^o$C/min. While the Mo source and SiO$_2$ growth substrate reach 750 $^o$C, the maximum temperature of the sulphur is $\approx$ 150 $^o$C. After a 19 min growth period, the furnace is opened, and the sample is rapidly cooled to room temperature in 500 sccm flowing Ar. A detailed description of the process and characterization of typical samples can be found in refs. \cite{Pollmann.2018,Pollmann.2020}
\\
\\
\noindent
\textbf{MoS$_2$ transfer}\\
The transfer of MoS$_2$ to the Au substrates was accomplished by spin-coating PMMA onto the as-grown sample, which was then placed in a bath of 0.1 M KOH to slowly etch the SiO$_2$ surface and release the PMMA/MoS$_2$ layer from the substrate. The floating PMMA/MoS$_2$ layer was then transferred into successive water baths for cleaning and finally scooped onto the target substrate. The sample was then dried, and the PMMA was removed with an acetone spray, followed by an acetone bath.
\\
\\
\textbf{Irradiation and ToF-SNMS} \\
Highly charged xenon ions were generated in an electron beam ion source (EBIS) commercially available from Dreebit GmbH, Germany. To select and modify the beam parameters (see supporting information for detailed beam characterization) a bending sector magnet and an deceleration section described in ref. \cite{Skopinski.2021} has been used. Due to ion irradiation sputtered sample particles are post ionized by an Excimer laser (ExciStar XS by Coherent) with a wavelength of 157 nm. The laser beam is shot parallel to the sample surface with a distance of 1.2 mm and has a beam diameter of 0.21 mm. The ionized particles are extracted into a time-of-flight mass spectrometer. We vary the extraction delay between time of HCI impact and extraction into the spectrometer and probe which particles have reached the laser volume in each case. From every mass spectra the $^{98}$Mo-peak was evaluated and is fitted in Fig \ref{sch:5keV_Au_sd} and \ref{sch:260keV_Au_sd2} normalized to the number of primary HCIs per pulse (see Electronic Supplementary Information for primary pulse characterization) as a function of the extraction delay. For the measurements with Xe$^{17+}$ and Xe$^{37+}$ at a kinetic energy of 5 keV as well as Xe$^{28+}$ at 260 keV the determined number of ions per pulse had to be corrected after the measurement series. The correction factor is indicated in the corresponding figure. The data shown in Fig \ref{sch:5keV_Au_sd}, \ref{sch:260keV_Au_sd2} and \ref{sch:comp-Au-vs-SiO2} has been smoothed by Lowess filters with OriginLab 2019b to reduce  noise. The description and figure S1 in ESI\dag~ gives an impression of the not smoothed data.
\\
\\
\textbf{MD simulations}\\
Empirical potential MD simulations using the LAMMPS package \cite{Plimpton.1995} are carried out to model 5 keV Xe ion impact on 2D-MoS$_2$ supported by gold substrate.
For MoS$_2$ either reactive bond order (REBO) potential \cite{Liang.2012} or the Stilinger-Weber (SW) potential \cite{Jiang.2013} are used to parametrize the interactions of Mo and S atoms -- the results turn out to be qualitatively comparable. The Au(111) substrate is modelled using the embedded atom method (EAM) potential.\cite{Ackland.1987} Van der Waals interactions of the 2D material and the substrate are taken into account by Lennard-Jones (LJ) potentials with the parameters fitted to PBE-vdW calculations given in table \ref{tbl:theo} below.
\begin{table} [h]
  \caption{\ Parameters of the Lennard-Jones interatomic potential.}
  \label{tbl:theo}  
  \begin{tabular}{lllll}
    \hline
       &    & eps & sigma & rcut \\
    \hline
     S   & Au  & 0.00448 & 3.264 & 9.3025\\
     Mo & Au & 0.00203 & 2.282 & 8.185  \\
    
  \end{tabular}
\end{table}\\
The interaction with the high energetic ion and interactions at close seperations are parameterized by the Ziegler-Biersack-Littmark (ZBL) potential \cite{Ziegler.1985} (smooth joining for REBO, SW). 
The considered model system consist of the 2D material on a 3.5 nm thick Au(111) substrate. The lateral extension of the simulation box amounts to 18.3 $\times$ 15.8 nm², which results in simulations comprising 75 k atoms in the simulation box. As in previous studies the impact energy is dissipated by Berendsen thermostats at the boundaries of the simulation box.
During the MD run all atoms passing the top boundary region are accounted for together with their velocities.
The statistics of the Mo velocity distribution is collected from 1200 ion impacts distributed over 240 impact points which are selected in the irreducible area.

\section{Results and discussion}
For our study we have chosen monolayers of MoS$_2$, as this is the best investigated 2D material apart from graphene. In contrast to graphene, its constituent Mo can be unambiguously detected by our mass spectrometer and the material is thus well suited for our purposes. The experimental set-up has been described in detail elsewhere, \cite{Skopinski.2021} and here we will only briefly summarize its main capabilities. Our ultrahigh vacuum set-up is based on a reflectron type time-of-flight mass spectrometer, designed to analyse secondary particles emitted during ion irradiation. We use an electron beam ion source to produce bunches of ions with various charge states q (here between Xe$^{17+}$ and Xe$^{40+}$) and a deceleration/acceleration section to tune the kinetic energy of the ions, here 5 keV and 260 keV. In addition, our set-up is equipped with a pulsed excimer laser for post-ionization of the sputtered particles, since most of them are emitted as neutral particles. By delaying post-ionization and extraction with respect to the time of impact of the ion pulse on the sample -- corresponding to a measurement of time-of-flight mass spectra at different extraction delays -- we can obtain mass and velocity distributions of the emitted neutral particles. For the data that we will present in the following, from each spectrum the intensity or the $^{98}$Mo mass peak was evaluated and plotted as a function of the extraction delay. The higher the extraction delay of a signal, the lower the velocity at which the associated particle moves away from the sample surface. 
Note, that we obtain the velocity distribution from the flight time distribution and that for a quantitative analysis a very time-consuming optimization of the set-up would be required. For the purpose of this paper it is however sufficient to analyze the relative numbers and we therefore refrain from plotting absolute velocities except for the comparison with our simulations.

\subsection{Dependence on the potential energy}
We begin by presenting and discussing the data obtained in the experiments on irradiation of MoS$_2$ on Au (see methods section for preparation) with Xe$^{q+}$ ions at a low kinetic energy of 5 keV. In this case we expect some energy transfer to the sample via nuclear collisions $E_{kin}^{nuclear}$ (q) while there should be almost no electronic contribution $E_{kin}^{electronic}$ (q) due to the low velocity of the ion. By varying the charge state and therefore the potential energy of the ion we change the deposited energy $E_{dep}^{pot}$ (q) which is known to be deposited within the first few layers of the sample material.\cite{Arnau.1997,Aumayr.2004}

\begin{figure}
\centering
  \includegraphics[width=1\linewidth]{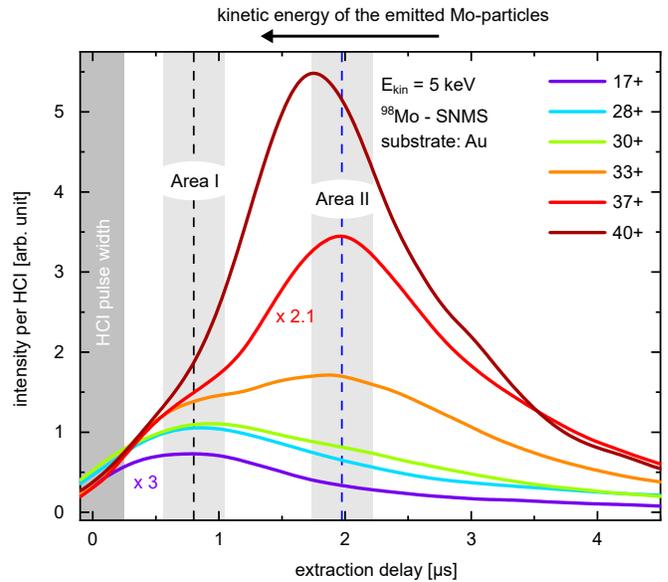}
  \caption{Intensity of the $^{98}$Mo signal as a function of the extraction delay for different charge states. A correction factor of 3 (Xe$^{17+}$) and 2.1 (Xe$^{37+}$) has been used to account for HCI-current calibration difficulties. The marked grey areas depict contributions of kinetic (area I) and potential (area II) energy to the emission process. The dark grey area shows the estimated pulse width of the HCI pulse.}
  \label{sch:5keV_Au_sd}
\end{figure}

Figure \ref{sch:5keV_Au_sd} shows the intensity of emitted neutral Mo particles during the irradiation per primary HCI as a function of the extraction delay of the corresponding mass spectra. The purple curve stands for the irradiation by 5 keV Xe$^{17+}$ ions, being equivalent to a potential energy of $E^{pot}$ = 3.0 keV, as the lowest charge state used. The measured data of the Mo signal shows one maximum at the extraction delay t$_{extraction}$ = (842.9 ± 10.5) ns. As the potential energy is significantly smaller than the kinetic energy, we assume that the latter has the greatest influence on the emission process. We therefore assign the higher velocity Mo particles in the region marked ``area I'' to the kinetic sputtering process originating from a linear collision cascade.\cite{Sigmund.1981} 

This obviously changes when the contribution of the potential energy increases. The measured signals change significantly with respect to the shape, intensity and position of the maximum for the charge states q between 28 and 40. Around the extraction delay of 1.975 µs, a second contribution to the signal develops, which strongly increases with a increasing charge state of the HCI. For the Xe$^{40+}$-irradiation with a potential energy of $E_{pot}$ = 38.5 keV, which is almost seven times greater than the kinetic energy (E$_{kin}$ = 5 keV), this additional contribution dominates the spectrum with its maximum at t$_{extraction}$ = (1684.1 ± 3.2) ns, marked as ``area II''. 

\begin{figure}
  \includegraphics[width=0.9\linewidth]{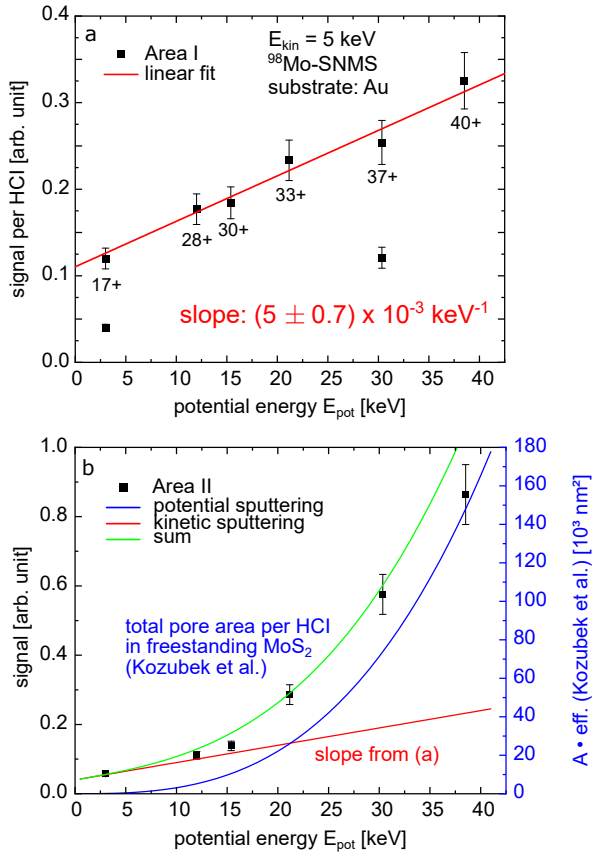}
  \caption{Evaluation of area I (a) and II (b) in Fig.~ \ref{sch:5keV_Au_sd} as a function of the potential energy of the projectile. The data in (a) can be described by a linear fit of slope m = 5 $\times10^{-3}$~keV$^{-1}$. Comparison of signal due to potential sputtering (area II, (b)) with the results for potential sputtering of MoS$_{2}$ by Kozubek et al.\cite{Kozubek.2019} The green curve is the sum of potential sputtering (blue) and kinetic sputtering (red; a linear dependence with the same slope as in (a) was used here). }
  \label{sch:5keV_Au_sd2}
\end{figure}

Overall, we observe that an increase in the potential energy of the HCI, and therefore the electronic excitation energy $E_{dep}^{pot}$ (q) of the sample, clearly enhances the total yield of emitted Mo-particles. This effect has been found for 3D materials such as LiF and has been coined {\it potential sputtering}.\cite{Aumayr.2004} That we can identify a similar process here in a 2D material, underlines the fact that the energy deposition is indeed limited to the very first few layers. 
In addition, we find that the increase in number of sputtered particles is due to a significant contribution of particles with a lower emission velocity (``area II'') than those assigned to a kinetic sputter mechanisms (``area I''). This has never been observed before, and we will discuss this important finding in more detail in the following paragraph.
Since the transmission through our spectrometer is energy dependent, we selected two areas with sputtered particles of a different energy and only relate intensities from the same area. In Fig.~\ref{sch:5keV_Au_sd2} we plot the signal at the fixed emission energies (area I and II) as a function of the potential energy of the HCI (corresponding to its charge state). 

From Fig.~\ref{sch:5keV_Au_sd2} it is obvious, that particle emission in the two areas exhibits a different dependence on the amount of deposited potential energy. In area I (0.8 µs), we observe a linear dependence of the signal on the potential energy of the projectile. The fit shown in Fig.~\ref{sch:5keV_Au_sd2}a has a slope of ($5\pm0.7)\times10^{-3}$~keV$^{-1}$.
In contrast, the signal in Area II (1.975 µs) increases more strongly with the potential energy of the primary HCI. Many previous studies on the interaction of HCIs with surfaces identified the so-called threshold values for the creation of defects (for a detailed overview see Aumayr et al.).\cite{Aumayr.2011} Our data indicates that potential sputtering can be detected even for the smallest charge states we have used and becomes the dominant mechanism for the emission of slow particles by projectiles with a potential energy of more than $E_{dep}^{pot}$~(q=30)~=~17~keV. 

After having observed the transition from and to kinetic and potential sputtering mechanisms in the emission process, the question arises, if we can unravel their interplay and interdependency, respectively. If, e.g., there was sufficient potential energy deposited into the 2D material and the resulting sputtering process would be fast enough, all of the sample material will have been emitted, before the emission on the basis of the linear cascade sets in. That is, a linear cascade can still take place in the substrate, but there would no longer be any sample material left to be affected. We observe from our data, that even for the highest charge state, Xe$^{40+}$, the signal in area I (identified as originating from kinetic sputtering) remains more or less unchanged. The small increase in area I with increasing potential energy is most likely due to a widening of the signal from area II upon increase. We thus deduce that the deposition of potential energy has no significant effect on the processes related to kinetic sputtering. There are several possible reasons for this, e.g.: (i) the excitation of the target material by the HCI is in general too weak, (ii) the excitation has already decayed, or (iii) is still building up. While electronic excitation is clearly strong enough to sputter particle, we will consider the different time scales of proposed models to describe the interaction of HCIs with solids in the following section.

Next, we will discuss the super-linear behaviour observed in area II. In Fig \ref{sch:5keV_Au_sd2}b we used a combination of the linear fit determined by fitting the data in Fig.~\ref{sch:5keV_Au_sd2}a and the dependence of pore radius in MoS$_2$ as a function of potential energy as determined by Kozubek et al. In this work we used high resolution scanning transmission electron microscopy to analyze the pore formation in freestanding MoS$_{2}$ irradiated by HCI and revealed a linear dependence of both, the pore radius and efficiency of pore formation, on the potential energy of the HCI. \cite{Kozubek.2019} The total pore area as a product of the pore area and efficiency is shown in Fig.~\ref{sch:5keV_Au_sd2}b in blue. Since kinetic sputtering processes are strongly reduced in freestanding 2D samples, we assume that the determined dependence represents exclusively the potential effects. For our sample of substrate supported MoS$_{2}$, we clearly have to account for the substrate driven kinetic sputtering. The linear fit in Fig.~\ref{sch:5keV_Au_sd2}a that has been transferred to Fig.~\ref{sch:5keV_Au_sd2}b, is an attempt to account for exactly this contribution. The sum of kinetic sputtering due to the substrate and potential sputtering measured for a freestanding sample matches the data very well. We find no indications for synergistic effects. We therefore conclude that for the sputtering of supported 2D MoS$_{2}$ by slow HCIs, effects from potential sputtering simply add up to the ones from kinetic sputtering. 

Several possible mechanisms for potential sputtering have been suggested. Many of them have originally been applied to describe the electronic excitation of solids via swift heavy ions. In this case, the projectiles are so fast, that the scattering cross section for nuclear collisions is practically zero and the slowing down of particles (stopping) happens via electronic excitations and ionization of target atoms. This is very different from HCI, but the dominant electronic excitation and similar irradiation induced morphologies \cite{Aumayr.2011} have led researchers to use the models anyhow. 

One possible scenario, the so-called “Coulomb explosion” is a burst of material due to Coulomb repulsion of the positively-charged atoms around the point of impact where electrons have been depleted by the HCI. Multiply-charged secondary ions or clusters could thus be an indication of such a Coulomb explosion.\cite{Bitensky.1987} Further, mechanistic arguments as well as model calculations predict the emission of rather energetic particles,\cite{Cheng.1997} which is in clear contrast to our observations.

Defect-mediated desorption is observed in materials where a self-trapped exciton (STE) forms upon irradiation like in alkali halides or in SiO$_2$. In our case, this mechanism seems highly unlikely, as the Au substrate would efficiently quench any excitonic processes (typically no PL signal is observed from these samples).\cite{Pollmann.2020}

Another possibility is the so-called non-thermal melting, describing the destabilization of atomic bonds caused by the direct promotion of electrons from bonding valence band into anti-bonding conduction band states.\cite{Schenkel.1998} This scenario has been intensively discussed to explain laser ablation.\cite{Stoian.2000} In this case, two mechanisms are relevant, again Coulomb explosion as a consequence of the electron emission due to the strong electronic excitation, and the direct emission of atoms from an electronically repulsive state with extremely low, sub-thermal energies on a sub-ps time-scale. If the latter was dominant, we would expect that the effective contribution from kinetic sputtering (see below) should decrease with increasing potential energy as atoms would be efficiently ejected before the collisional cascade sets in.

Finally, the thermal spike model assumes the dissipation of the electronic energy via energy transfer to the lattice atoms. If this local increase of the temperature rises above the melting point of the target, it may cause structural modifications or even the emission of particles \cite{Toulemonde.1992,Osmani.2011,Madau.2017} similar to thermal emission from a hot target. The velocity distributions obtained in our experiments indicate a large fraction of particles emitted at energies comparable to thermal emission, which increases with increasing charge state.

\subsection{Dependence on the kinetic energy}
\begin{figure}
  \includegraphics[width=1\linewidth]{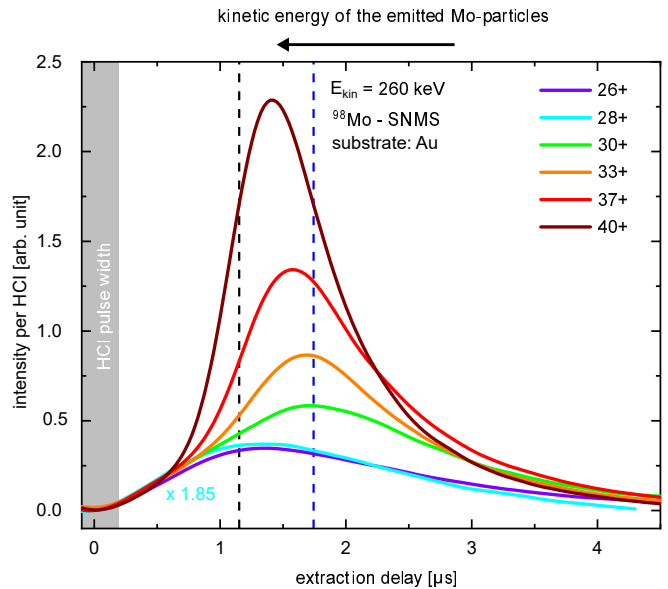}
  \caption{Intensity of $^{98}$Mo-particles emitted from a monolayer MoS$_2$/Au sample during irradiation with 260 keV HCIs as a function of the extraction delay. A correction factor of 1.85 has been used for the Xe$^{28+}$ measurement to account for current calibration issues. The half maximum width of the widest primary ion pulse used in this experiment t(Xe$^{33+}$)~=~397.3 ns is highlighted in dark grey at the extraction delay t$_{extraction}$~=~0~µs. The two areas of particular interest are marked by black an blue coloured lines.}
  \label{sch:260keV_Au_sd2}
\end{figure}

In the following section we will present and discuss the case where the kinetic energy plays a more dominant role. To this end, we repeated the experiment with faster 260 keV HCI with the same charge states. As presented in Fig.~\ref{sch:260keV_Au_sd2}, we observe a shift of the highest intensity towards later extraction delays for the lowest charge state; but again, a strong increase in the intensity with increasing potential energy of the HCI. 
We start our analysis again with the lowest charge state (Xe$^{26+}$, E$_{pot}$ = 9 keV) as the influence of the kinetic energy (E$_{kin}$=260 keV) can be best observed there. The position of the maximum (black dashed line in Fig.~\ref{sch:260keV_Au_sd2}) is with 1.35 µs about 0.55 µs later than for the 5 keV measurement. With a similar charge state, the maximum is shifted towards particles emitted at lower velocities. Therefore, the starting position is already changed. Due to the increased kinetic energy, area I can no longer be assigned to the linear cascade regime, but must be attributed to a collisional spike. 

Those kinetic interactions are well studied, e.g. Seah showed that the sputter yield of gold by irradiation with Xe projectiles for velocities over 100~keV can be described by adding a spike contribution.\cite{Seah.2007} Fig.~\ref{sch:260keV_Au_sd2} also shows that the increase in potential energy clearly leads to an intensity increase. For Xe$^{30+}$, it starts at around 1750 ns (blue dashed line) with almost the same extraction delay as area II for the Xe 5~keV measurement series. For even higher charge states, the maximum shifts further towards smaller extraction delays, i.e.~that while the potential energy still leads to an additional contribution by slower particles, their velocity seems to increase slightly with the charge state of the projectile. This could be due to a synergy effect as the thermal spike overlaps with the collisional spike in time and in space.

\subsection{Dependence on the substrate}
\begin{figure}
  \includegraphics[width=1\linewidth]{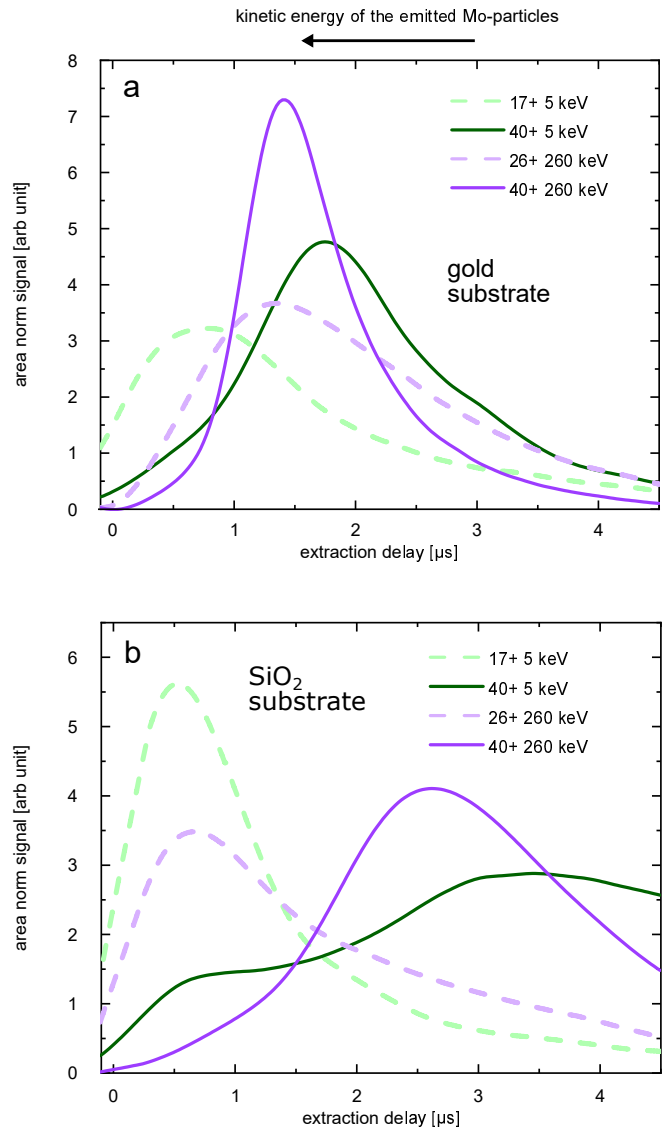}
  \caption{Area-normalized (0 to 6 µs) intensity of $^{98}$Mo-particles emitted from a) a SL MoS$_{2}$/Au sample and b) a SL MoS$_{2}$/SiO$_{2}$ sample during bombardment with Xe$_{5 keV}^{17+}$ - (dashed green), Xe$_{5~keV}^{40+}$ - (green), Xe$_{260~ keV}^{26+}$ - (dashed purple), and Xe$_{260~keV}^{40+}$ ions (purple). Plotting against the extraction delay allows slower and faster particles to be distinguished.}
  \label{sch:comp-Au-vs-SiO2}
\end{figure}

To further understand the dependence on the kinetic energy of the projectile, we compared the Mo particles sputtered from single layer MoS$_{2}$ transferred onto two different substrates, i.e., gold and silicon dioxide. For each system, the smallest and largest charge state used for E$_{kin}$ = 5~keV and E$_{kin}$ = 260~keV, are summarized in Fig.~\ref{sch:comp-Au-vs-SiO2}. For better comparability and to eliminate the sources of error in the primary ion normalization, Fig.~\ref{sch:comp-Au-vs-SiO2} shows the smoothed and area-normalized (0 to 6 $\mu$s) SNMS data sets for the Au substrate a) and the SiO$_{2}$ substrate b). Dashed lines represent the smallest and solid lines the highest charge states. Please note, that while we will focus in the following on the shift in the contribution of slower Mo particles, we want to emphasize that an increase in the energy of the projectile, both kinetic and especially potential, always results in an increase of the Mo yield. This agrees perfectly well with our earlier results on the interaction of HCIs with freestanding MoS$_{2}$ \cite{Kozubek.2019} and with supported hBN.\cite{Kozubek.2018}

Fig.~\ref{sch:comp-Au-vs-SiO2}b shows that for both kinetic energies the position of the maximum differs significantly for the smallest and highest charge state. For the charge states Xe$_{5 keV}^{17+}$ and Xe$_{260 keV}^{26+}$, the maximum is at an earlier extraction delay ($<1$ µs) than for Xe$_{5 keV}^{40+}$ and Xe$_{260 keV}^{40+}$ ($> 2$ µs). 

In the previous discussion, the positions of the maxima  in the velocity distributions were either found in area I or area II. Area I was assigned to the kinetic energy dominating the emission process and area II to the potential energy being most relevant. As already discussed, for the Au substrate (Fig.~\ref{sch:comp-Au-vs-SiO2}a), this clear distinction works well for the Xe 5 keV measurement series. The maxima of the Xe 260 keV measurement series, on the other hand, are found almost at the same extraction delay, which we attributed to the occurrence of a collisional spike in the gold substrate at this kinetic energy. In contrast, for a SiO$_2$ substrate (Fig.~\ref{sch:comp-Au-vs-SiO2}b), the data shows clearly distinct positions of maxima for the both charge states at both kinetic energies. From our discussion above, we would thus infer that in SiO$_2$ there is no indication for a collisional spike. Obviously, the nature of the emission processes therefore does not only depend on the kinetic energy but also on the substrate underneath the 2D sample. 

Due to the nature of a 2D material this is not so surprising and in agreement with earlier findings. Theoretical simulations show that the substrate can indeed be the key to the modification of a supported 2D material due to sputtering by back-scattered ions, sputtered substrate atoms,\cite{Kretschmer.2018} or by e.g. a change of strain.\cite{Krasheninnikov.2020} To compare the two substrates, we carried out SRIM calculations \cite{Ziegler.2010} for 260 keV Xe ions in a monolayer of MoS$_{2}$ on a substrate. We find a nuclear stopping power of 5.7 keV/nm for a gold substrate and of 2.59 keV/nm for a SiO$_{2}$ substrate.
The higher efficiency for the transfer of kinetic energy from the ion to the substrate lattice in case of gold makes it reasonable that we observe the signature of a collisional spike for the gold substrate but not the for the SiO$_{2}$. Remarkably however, we observe no indication that the contribution from potential sputtering of the monolayer MoS$_{2}$ is affected by the substrate. In principle, the response of the two substrates, one being a metal and the other an insulator, towards irradiation with HCIs should be fundamentally different.\cite{Aumayr.2004} Part of the explanation might be that due to the preparation procedure used here, a layer of intercalated water might be present.\cite{Ochedowski.2014, Pollmann.2020} It could act as a protective coating between the 2D material and its substrate. Nevertheless, we believe the main reason to be the extremely shallow energy deposition, which was just recently again demonstrated by Schwestka et al.~, where pore creation via HCI irradiation was achieved with atomic depth precision.\cite{Schwestka.2020} 

\begin{figure}
\includegraphics[width=1\linewidth]{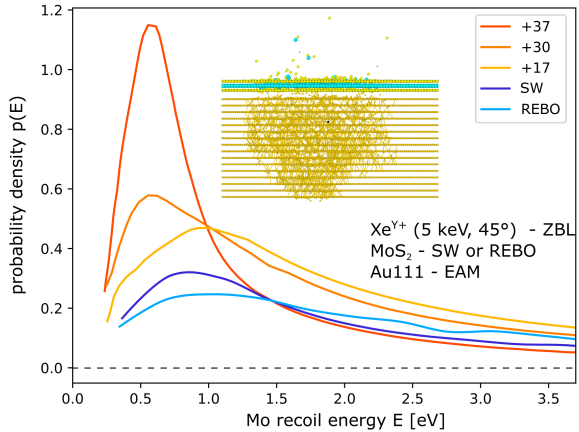}
  \caption{Probability density of the Mo recoil energy distribution for 5 keV Xe ion impact on 2D-MoS$_2$ supported by Au(111) substrate. The simulations results for two different interaction potentials (SW, REBO) of MoS$_2$ modelling neutral projectiles are compared to the experimental results for charged Xe (with different charge states). The inset shows a typical atomic configuration just after the ion impact. Cyan balls represent Mo atoms, yellow sulfur and orange Au atoms.}
  \label{sch:comp-exp-MD}
\end{figure}
In order to better understand the experimental finding of the kinetic energy driven sputtering mechanism at low charge states we carried out molecular dynamics (MD) simulations of 5 keV Xe ion impacts on MoS$_2$ supported by Au(111) substrate.
Empirical potential MD accounts for the nuclear stopping of an neutral projectile only and can therefore be used to view the limiting case of neutral projectile impinging on the system of interest.
A detailed description of the simulation setup can be found in the method section. Fig. \ref{sch:comp-exp-MD} displays the extracted kinetic energy distribution of the sputtered Mo atoms in comparison with the experimental data from Figure \ref{sch:5keV_Au_sd} transformed into a velocity distribution.
The simulations were carried out with two different interaction potentials for the MoS$_2$ sheet, such that strong dependence on the empirical potential can be ruled out.
The simulations results show a much broader distribution at higher recoil energies compared to the experimental data for higher charge states. This supports the picture of ejecting fast Mo atom from ballistic collisions for low charge state and high kinetic energy. With increasing charge state the maximum value of the recoil distribution growths and shifts to lower values -- the higher the charge state the more low energy Mo recoils are measured -- and desintegration of the MoS$_2$ sheet is dominated by potential sputtering.

\section{Conclusions}
To understand the mechanisms of material modification under impacts of HCIs, we studied the distribution of sputterd Mo particles from a MoS$_2$ sheet on various substrates. To do so, we have implemented a suitable experimental approach and could demonstrate that velocity distributions of particles sputtered from 2D materials can be used to differentiate between various sputtering mechanisms. From our data we find that the potential energy of a HCI leads to the emission of slow neutral particles from the top layer of the sample. At low kinetic energies , i.e.~smaller or comparable to the potential energy, this contribution increases continuously with increasing potential energy. In this regime, there are no significant contributions from either very fast or from ultra-slow particles, leading us to rule out Coulomb explosion or non-thermal melting to be the dominant mechanisms. Both take place on ultrashort time-scales and would thus lead to particle emission before the collision cascade due to kinetic processes reaches the surface again. However, our sample material is ultrathin and and if a significant amount of material had been emitted, we would no longer be able to detect the signature of the slower processes.

The thermal energies of the emitted atoms point instead towards a mechanism that transfers potential energy from the HCI to the lattice predominantly via electron-phonon-coupling. At higher kinetic energies, i.e.~a factor of five or more in comparison to the potential energy, this fraction can still be identified, but is complemented by a second contribution, which in the case of gold we attribute to the ion induced collisional spike. The two mechanisms are clearly separable and appear to occur independently of each other. These experimental findings are consistent with the results of our atomistic simulations. While our data seems to be in accordance with the corresponding effects simply adding up, synergistic effects are nevertheless still possible, as both mechanisms occur on similar time- and length scales. 

In conclusion, we have implemented a powerful method which for the first time allowed us to investigate the velocity distributions of particles emitted from a 2D material upon HCI irradiation. Our approach not only provided new insights into the interaction between ions and solids, but also paves the way to a deeper understanding with respect to defect engineering.

\section*{Conflicts of interest}

There are no conflicts to declare.

\section*{Acknowledgements}
The authors acknowledge the German Research Foundation (DFG) by funding through projects SCHL 384/20-1 (Project Number 406129719), C05 (Project number 278162697) within the SFB1242 “Non-Equilibrium Dynamics of Condensed Matter in the Time Domain” and KR 4866/2-1.

\bibliography{HCI-sputter-paper} 

\end{document}